\documentclass[%
letterpaper,
reprint,
superscriptaddress,
amsmath,amssymb,
aps,
pra,
dvipsnames,svgnames,x11names
]{revtex4-2}

\usepackage{graphicx}
\usepackage{dcolumn}
\usepackage{bm}
\usepackage{amsmath,amssymb,graphicx,xcolor,units,amsfonts}
\usepackage{color}
\usepackage{verbatim}
\usepackage{tikz}
\usepackage{siunitx}
\usepackage{titlesec}
\usepackage{titletoc}
\usepackage{mathtools}
\usepackage{physics}
\usepackage{bm}
\usepackage{gensymb}
\usepackage{CJK}

\titleformat{\chapter}{\normalfont\huge}{\thechapter.}{20pt}{\bfseries\huge}
\usepackage[utf8]{inputenc}
\usepackage[title]{appendix} 

\usepackage{mwe}
\usepackage{subfig}
\usepackage{caption}
\usepackage{braket}
\usepackage[english]{babel}

\begin{document}

\preprint{APS/123-QED}
\date{\today}
\begin{abstract}
Miniature atomic beams can provide new functionalities for atom based sensing instruments such as atomic clocks and interferometers.  We recently demonstrated a planar silicon device for generating well-collimated thermal atomic beams \cite{Li2019}. Here, we present a near-source fluorescence spectroscopy (NSFS) technique that can fully characterize such miniature beams even when measured only a few millimeters from the nozzle exit.  We also present a recipe for predicting the fluorescence spectrum, and therefore, the source angular distribution, even under conditions of strong laser saturation of the probing transition.  Monte Carlo simulations together with multi-level master equation calculations fully account for the influence of optical pumping and spatial extension of the Gaussian laser beam.  A notable consequence of this work is the agreement between theory and experimental data that has allowed fine details of the angular distribution of the collimator to be resolved over 3 decades of dynamic range of atomic beam output flux.
\end{abstract}

                              
\begin{CJK*}{UTF8}{} 
\CJKfamily{gbsn}
\title{Near source fluorescence spectroscopy \\
for miniaturized thermal atomic beams}
\author{Chao Li (李超)}
\email{lichao@gatech.edu}
\author{Bochao Wei (魏博超)}
\author{Xiao Chai (柴啸)}
\affiliation{School of Physics, Georgia Institute of Technology, 837 State St, Atlanta, Georgia 30332, USA}
\author{Jeremy Yang}
\author{Anosh Daruwalla}
\author{Farrokh Ayazi}
\affiliation{School of Electrical and Computer Engineering, Georgia Institute of Technology, 777 Atlantic Drive NW, Atlanta, Georgia 30332, USA}
\author{C. Raman}
\affiliation{School of Physics, Georgia Institute of Technology, 837 State St, Atlanta, Georgia 30332, USA}
\maketitle
\end{CJK*}

\tableofcontents

\section{Introduction}
Microfabrication techniques have a remarkable opportunity to transform atomic sensors, normally laboratory scale devices, into portable instruments.  Such instruments are urgently needed for precise navigation and timing, electromagnetic field sensing and gravimetry, all applications where atoms provide a basic reference standard that is traceable to fundamental constants \cite{battelier2016development,mccarthy2018time,boto2018moving,bidel2018absolute}.  Prototypical examples of atomic platforms currently targeted for these applications are ultracold atoms on a chip and MEMS-based microfabricated alkali vapor cells \cite{Keil2016,Kitching2018}. 

In previous work, we demonstrated a continuous, miniature thermal atomic beam source that propagated along a silicon surface in microchannel arrays defined by photolithography \cite{Li2019}. Such well-collimated sources could find application in miniature atom interferometry \cite{gustavson2000precision} and enable precise studies of atom-surface interactions \cite{perreault2005observation}. Moreover, unlike vapor cells, they do not suffer from Doppler broadening or wall collisions \cite{happer1972optical}.  Thus, they combine the simplicity of thermal vapors with the spectroscopic purity of cold atom systems.  Since these sources are necessarily miniature, however, one needs to characterize the atomic beam close to the source itself, where the atoms' spatial and velocity distributions are mixed together.  This is in contrast to typical laboratory-scale atomic beam experiments operating in the far-field, where the two distributions can be separated \cite{citron1977experimental,keith1991interferometer,gustavson2000precision,schioppo2012compact,zheng2017measurement}.  In this work, we use fluorescence spectroscopy to perform a four-dimensional tomography, reconstructing the spatial and speed distribution for the chip scale atomic beam. We term this near-source fluorescence spectroscopy (NSFS).

\begin{figure*}[t]
\begin{minipage}{0.25\linewidth}
\centering
\subfloat[]{\label{fig:oven}\includegraphics[width = 0.9\textwidth]{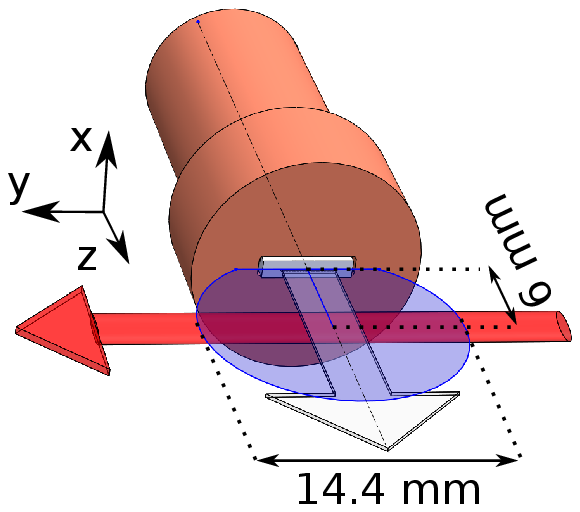}}
\end{minipage}
\begin{minipage}{0.35\linewidth}
\centering
\subfloat[]{\label{fig:ord_chip}\includegraphics[width = 0.9\textwidth]{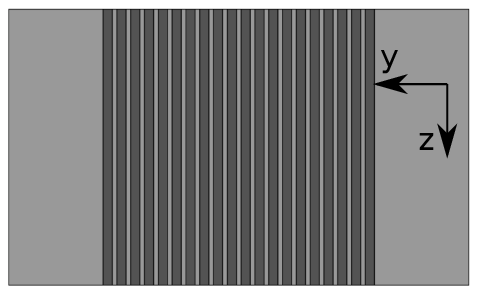}}
\end{minipage}
\begin{minipage}{0.35\linewidth}
\centering
\subfloat[]{\label{fig:cas_chip}\includegraphics[width = 0.9\textwidth]{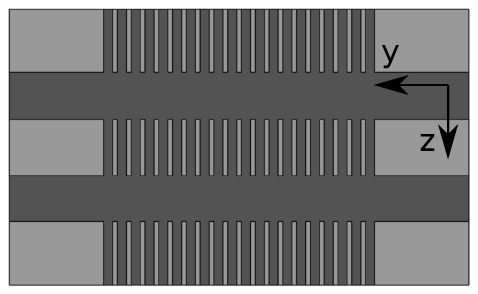}}
\end{minipage}\par\medskip
\begin{minipage}{.6\linewidth}
\centering
\subfloat[]{\label{fig:rsc}\includegraphics[width = 0.8\textwidth]{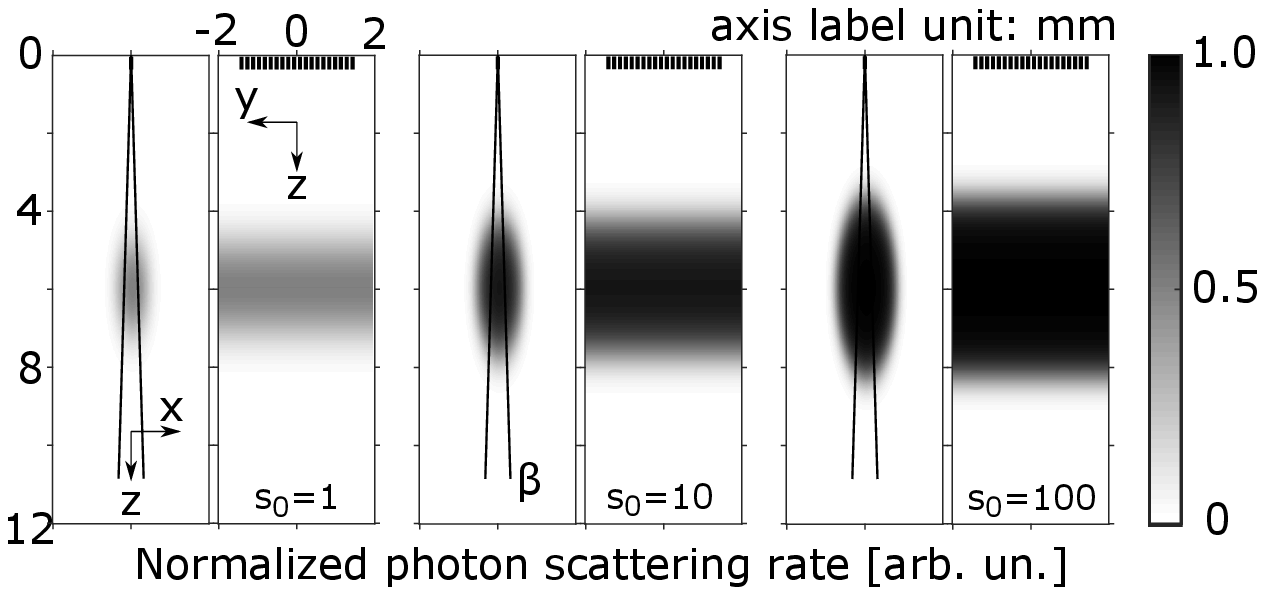}}
\end{minipage}\par\medskip
\captionsetup{justification=centerlast}	
\caption{
Miniature, microfabricated atomic beams probed by NSFS (near-source fluorescence spectroscopy).  
(a) is a cartoon image showing the close-up view of the front end for a compact atomic beam production and collimation apparatus. A microfabricated silicon device with 20 collimating channels sits inside a slit at the center of the copper head. The red arrow along $+\hat{y}$ represents the laser beam and the gray arrow along $+\hat{z}$ represents the atomic beams. The shaded region in blue shows the field of view of our fluorescence collecting system. 
(b) shows the cartoon top-view of such a collimator (termed an ordinary collimator, in contrast to the cascaded collimator of (c)) before bonding the sealing wafer on top. The chip is $3$ $\si{mm}$ $\times$ $5$ $\si{\mm}$ in size with 0.5 $\si{\mm}$ thickness. For an individual channel, $l/d=30$ ($l=3$ $\si{\mm}$ and $d=100$ $\mu$m). 
(c) shows the top-view of a 3-stage cascaded collimator ($\sim 660$ $\mu$m/stage and $\sim500$ $\mu$m/gap) with the same overall length and diameter for an individual channel before wafer bonding. 
(d) Estimated spatial distribution of the normalized resonant photon scattering rate, $s/(1+s)$ in terms of probe saturation parameter $s = I(\vec{r})/I_{sat}$, on planes defined by $y=0$ (side view) and $x=0$ (top view), for three different laser intensities corresponding to $s_{max}$ (or $s_{0}$) = 1, 10, 100. 
Each panel is $4\times12$ $\si{\mm}$. The laser beam has Gaussian radius (for $1/e^2$ intensity) $w_x =0.5$ $\si{\mm}$ and $w_z=1.4$ $\si{\mm}$. Bars at the top represent the array of collimator outputs. Two lines forming an angle $\beta$ as a guide depicts atoms emitted within the collimator full-width at half-maximum (FWHM) angle $\beta =2\theta_{1/2}=3.6 \degree$ \cite{giordmaine1960molecular}. 
}
\label{fig:oven_chip_rsc}
\end{figure*}
\section{Experimentation}

The fabrication procedure for the chip scale atomic collimator as well as the fluorescence measurement protocols have been described in detail in the previous work \cite{Li2019}.  We briefly review these here, showing the major components of the NSFS measurement in Fig.~\ref{fig:oven}-\ref{fig:cas_chip}. Thermal atomic beams generated by 20 silicon microcapillaries attached to an effusive oven were probed a few millimeters after the nozzle exit. 20 atomic beams spaced by 150 $\mu$m center-to-center distance, co-propagate along the $+\hat{z}$ direction on the $y-O-z$ plane as defined in Fig.~\ref{fig:oven_chip_rsc}. An external cavity diode laser at 780 nm used as a fluorescence probe is scanned over a 1 GHz range across the ${}^{87}$Rb D$_2$ $F=2$ to $F'=3$ transition at a rate of 5 Hz. The laser beam is propagating perpendicular to the traveling direction of the atom beam along the $+\hat{y}$ direction centered at $z_{0} \approx 6$ $\si{\mm}$ and it is linearly polarized along the $z-$axis to maximize the collected fluorescence. Fluorescence emitted from the volume shown in Fig.~\ref{fig:oven} at the intersection of the laser and atomic beams is collected through two 2-inch lenses (not shown) located $\simeq $3 inches above this volume.  Light is collected onto a photodiode and the photocurrent is amplified by a low noise current amplifier. Because of the transverse Doppler effect, the fluorescence collected at different laser detunings is sensitive to the transverse speed distribution of the atomic beam, which contains information about the collimating performance of our microfabricated silicon microcapillary array--better collimation means narrower transverse speed distribution. Part of the laser output is injected into a rubidium vapor cell at room temperature for saturated absorption spectroscopy, calibrating its operating frequency and assisting the scan control. 

We performed two sets of experiments for the two different types of  collimators shown in Fig.~\ref{fig:ord_chip} and \ref{fig:cas_chip}. The cascaded collimator (\ref{fig:cas_chip}) generates 40 times purer atomic beams compared to the ordinary type single-stage collimator (\ref{fig:ord_chip}). This is because the two gaps efficiently release atoms whose trajectory deviates from the central axis, thus, behaving as a transverse velocity filter, as discussed in Ref. \cite{Li2019}. For each collimator, we recorded  fluorescence spectra over 10 different laser intensities adjusted by varying the probe laser power.  All other experimental parameters such as laser beam alignment, beam width and propagation direction were kept identical from one collimator to another, so that the spectra could be directly compared with one another.

\begin{figure*}[htbp]
\begin{minipage}{0.32\textwidth}
\centering
\subfloat[]{\label{fig:theory_3d}\includegraphics[width = 0.86\textwidth]{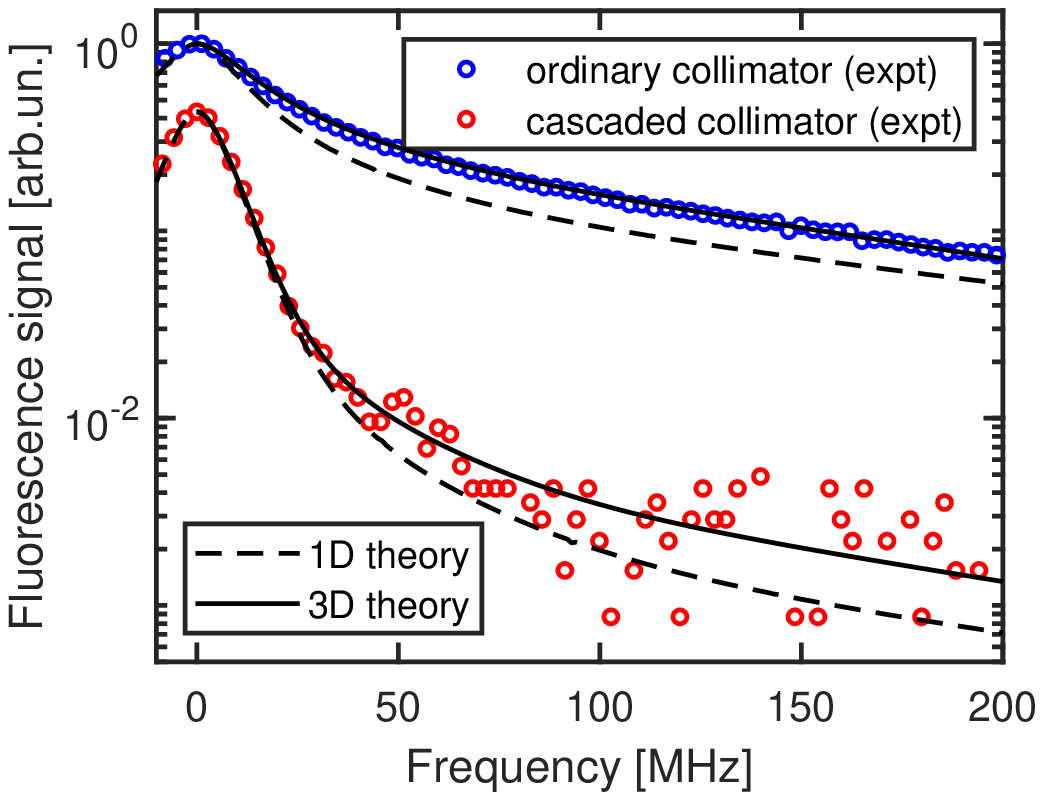}}
\end{minipage}
\begin{minipage}{0.32\textwidth}
\centering
\subfloat[]{\label{fig:pk_vs_s}\includegraphics[width = 0.95\textwidth]{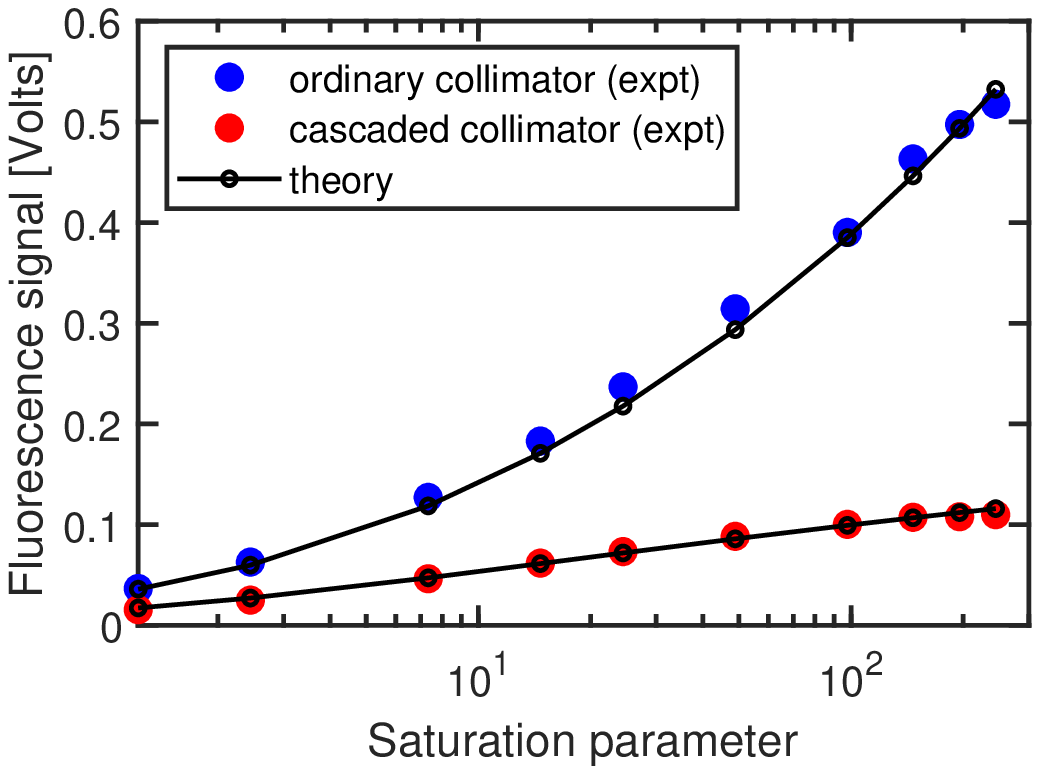}}
\end{minipage}
\begin{minipage}{0.32\textwidth}
\centering
\subfloat[]{\label{fig:hwhm_vs_s}\includegraphics[width = 0.88\textwidth]{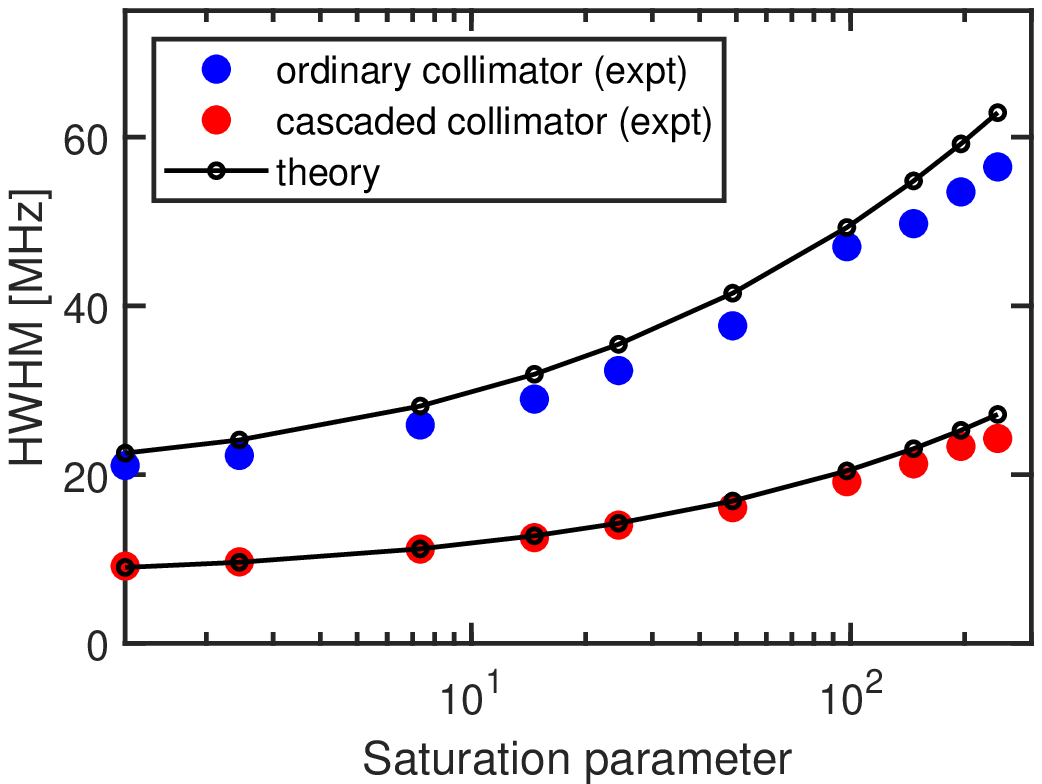}}
\end{minipage}
\begin{minipage}{0.32\textwidth}
\centering
\subfloat[]{\label{fig:hyperfine_levels}\includegraphics[width = 0.9\textwidth]{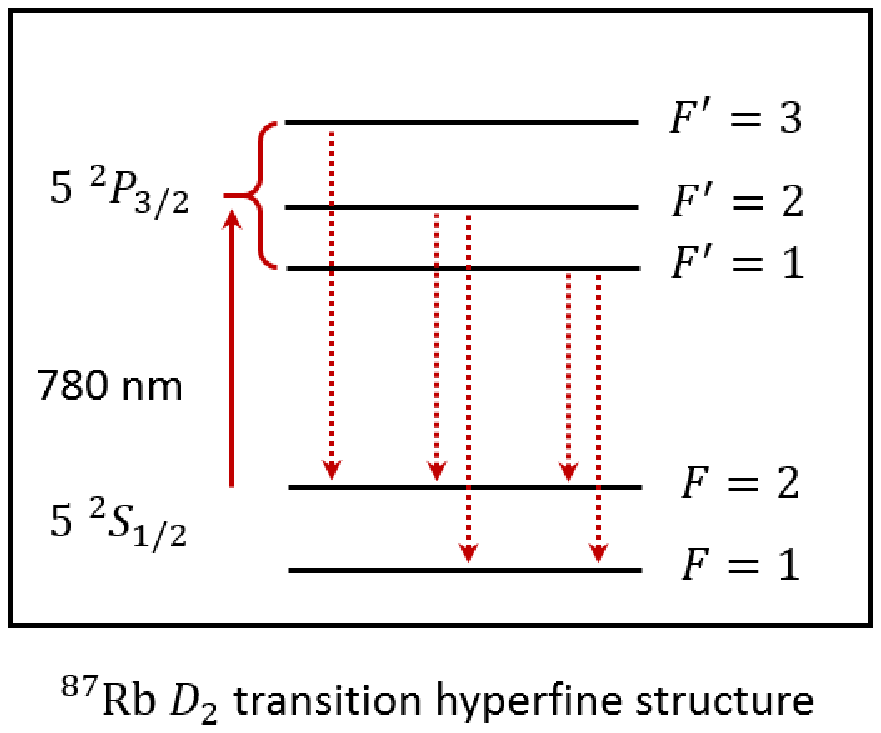}}
\end{minipage}
\begin{minipage}{0.32\textwidth}
\centering
\subfloat[]{\label{fig:dualspectra}\includegraphics[width = 0.9\textwidth]{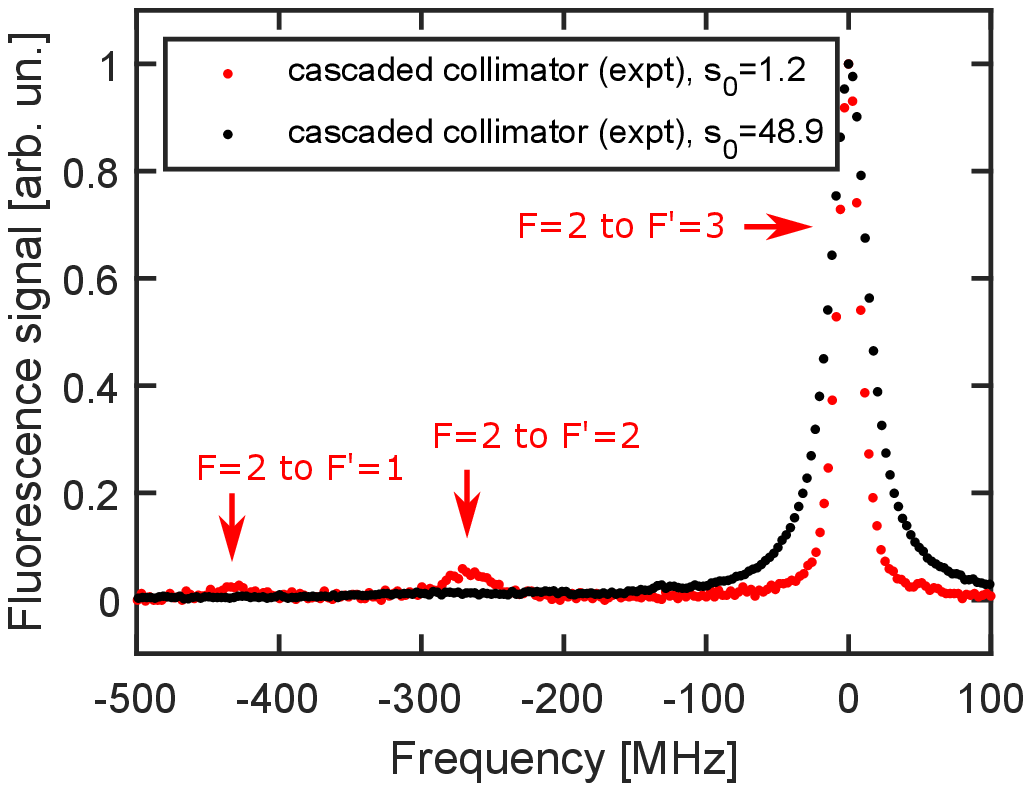}}
\end{minipage}
\captionsetup{justification=centerlast}	
\caption{
Measuring the angular distribution function of atomic collimators. 
(a) Fluorescence spectra vs. laser scan frequency at a saturation parameter of $s_0$=1.2 in a log scale. Blue (red) circles are experimental data for the ordinary (cascaded) collimator. 1D (3D) theory is shown in dashed (solid) lines.   
(b) Measured peak height of the fluorescence spectra vs. $s_0$. Blue (red) dots are experimental data for the ordinary (cascaded) collimator. Corresponding solid lines show the theory.
(c) Measured half width at half maximum (HWHM) of the fluorescence spectra vs. $s_0$. Corresponding solid lines show the theory. 
Error bars for the experimental data in (b) and (c) are smaller than the sizes of the blue and red dots. 
(d) Relevant ${}^{87}$Rb D$_2$ optical hyperfine transitions. $F$ labels the ground states. $F'$ labels the excited states. The solid curve with an arrow labels the laser excitation. Dotted lines with arrows label the spontaneous emission allowed by the selection rules. 
(e) Red (black) dots show the measured fluorescence spectra for the cascaded collimator at $s_0$=1.2 ($s_0$=48.9). Arrows indicate the hyperfine transitions. 
}
\label{fig:theory_exp} 
\end{figure*}
\section{Results}
\subsection{Experimental spectra}
In general, laser spectroscopy is most precise at low illumination (low saturation parameter $s = I/I_{sat}$), in order for the measurement to be minimally perturbative of the system under study.  Here $I_{sat}$ is the saturation intensity for the particular optical transition being probed.  However, fluorescence detection benefits from a higher probe intensity in order to overcome background noise caused by environmental light and detector/electronic noise sources.  Therefore, in practice one needs to work at an appreciable value of $s$.  In our case we would like to resolve the fluorescence spectrum at large detunings to probe the atomic emission at large angles to the collimator axis.  Therefore, we have developed a method to extract this velocity distribution even at finite values of $s$ where saturation cannot be ignored.  

Fig.~\ref{fig:theory_exp}a shows our main experimental result. We show experimentally measured spectra over a 200 MHz range of detunings that agree very precisely with a full three-dimensional numerical calculation according to Eq. (\ref{eq:fluorescence}). By contrast, the simple one-dimensional theory used in our earlier work \cite{Li2019} shows clear deviations at the level of 30$\%$ that are particularly pronounced for the ordinary collimator of Fig.~\ref{fig:ord_chip}.  While the best agreement occurs, as expected, for low saturation parameters, we have also systematically investigated in Fig.~\ref{fig:pk_vs_s} and \ref{fig:hwhm_vs_s} the influence of laser intensity on the fluorescence spectra up to saturation parameter $s_0=I_0/I_{sat} = 250$.  Here $I_0$ is the laser peak intensity and $I_{sat}=$ 3.05 mW/cm$^2$ is for linearly polarized light coupling the ${}^{87}$Rb D$_2$ $F=2$ to $F'=3$ transition.  Good agreement is found throughout the range of parameters explored.

In Fig.~\ref{fig:theory_exp}a we have plotted data on the blue side of the atomic resonance, as it avoids contamination from the $F=2$ to $F'=2, 1$ transitions occurring at -267 and -424 MHz, respectively, as well as contributions from the $^{85}$Rb isotope's hyperfine transitions \cite{SteckRb2015}. Therefore, we can regard our atoms as a pure two-level system for the $F=2$ to $F'=3$ transition, with the spectral wings truly representative of the atoms' transverse Doppler velocity distribution.  The transition natural linewidth is $\Gamma$ = $2\pi\times6.1$ MHz, but other mechanisms contribute to broadening the spectra.  These include power broadening, transit-time broadening, Doppler broadening, Zeeman broadening, collision broadening, as well as the finite laser linewidth \cite{foot2005atomic}. The power broadening dominates over these other mechanisms for the parameters of our experiment, and scales as $\Gamma\sqrt{1+s}$ with respect to the saturation parameter. For example, at 100 C, the transit time for atoms going though the laser beam is estimated to be $2w_z/\bar{v}$ $\approx$ 9 $\mu$s contributing a transit-time broadening about 0.1 MHz. The nozzle heater coil wrapped onto the copper head (shown in Fig.~\ref{fig:oven}) has a DC current running through it that generates a magnetic field with a strength less than 1 Gauss, corresponding to less than $\sim$ 0.23 MHz line broadening \cite{SteckRb2015}. No significant Zeeman broadening is also verified by momentarily turning off the nozzle heater and monitoring if the spectra get narrower. 
Since the mean free path is larger than the probe beam diameter, the collision broadening can be neglected because of the low atom number density of $10^8$ atoms/cm$^3$ 6 mm away from the nozzle exit at T = 100 C.
The laser linewidth is $<1$ MHz.  Hereafter, we discuss mainly power broadening and the Doppler effect, since all other mechanisms together contribute a total broadening at the $\sim$ 1 MHz level that can be neglected.

It is interesting to compare the different theoretical curves shown in Fig.~\ref{fig:theory_3d}. Our 1D theory assumes that the atom number density only varies with respect to the transverse coordinate $y$, and not the transverse coordinate $x$.  It is a useful approach if the laser interrogation occurs {\em far from} the source, i.e. $z_0 \gg 2w$, where $w$ is the Gaussian beam waist of the probe.  In this limit, one can neglect off-axis atomic trajectories along the $x$-directions, as well as laser intensity variations.  In reality, however, our probe is very close to the source, as seen in Fig.~\ref{fig:oven} and \ref{fig:rsc}, and one must account for the full three-dimensional nature of both the laser beam profile as well as atomic trajectories through the beam.  For our data, we see that the 1D theory accurately reproduces the spectral full-width at half-maximum (FWHM) but not the wings of the data.  Atomic trajectories at large angles to the central axis contribute most to these wings.

To gain a further qualitative understanding, we can divide the population into two components:  atoms traveling inside $\theta_{1/2}$ (beam component) and outside $\theta_{1/2}$ (diffuse component).  The half width of the transverse speed distribution for the beam component can be reasonably estimated by $2\bar{v}\theta_{1/2}$, while for the diffuse component it is much larger since atoms emitted into the diffuse component originated from atom-tube reflections within the source. We can think of the laser beam as a bundle of light rays traveling along $y$ with different $x$ and $z$ coordinates. All rays that lie in the $x=0$ plane intersect both the beam component and the diffuse component.  However, rays in the $z=z_0$ plane with $|x| > z_0\theta_{1/2}$ will interact only with the diffuse component.  Thus, the latter will measure a broader spectra. We quantify this result further in Fig.~\ref{fig:spectra_off_x}. The final spectrum will be a sum over rays with different coordinates, weighted by the contributions of different atomic densities, trajectories and laser intensities.  For our simulations, we assumed that the atomic speed distribution emitted into any direction obeys the Maxwell-Boltzmann distribution (in traditional atomic physics texts this is known as setting the deformation function to 1 \cite{Beijerinck1975,lucas2013atomic}). 

Armed with this theoretical approach, we apply it to a variety of different saturation parameters used in the experiment. In Fig.~\ref{fig:pk_vs_s} and \ref{fig:hwhm_vs_s}, we show the peak height and half width at half maximum (HWHM) vs. saturation parameter $s_0$ of values swept from 1 to 250 for two sets of measured fluorescence spectra for the ordinary collimator and the cascaded collimator, respectively. Our theory can successfully reproduce the measured peak fluorescence intensity as well as the HWHM of the measured spectra. This agreement was achieved without any adjustable parameters, assuming the known saturation parameter of $I_{sat}=$ 3.05 mW/cm$^2$ for linearly polarized light coupling the ${}^{87}$Rb D$_2$ $F=2$ to $F'=3$ transition. What appears to be peculiar about this result is that there is no apparent saturation effect at all. The measured curves in Fig.~\ref{fig:pk_vs_s} continue to increase with laser intensity, although the rate of growth at high values of $s$ is slower for the cascaded collimator than for the ordinary collimator.  This apparently counterintuitive behavior is a consequence of the laser interrogation region being close to the source, a condition not satisfied in most experiments, where a clear saturation can be observed \cite{citron1977experimental}.  As the laser intensity increases, the scattering rate in the center of the beam saturates, while in the wings, it can continue to grow.  The ordinary collimator contains a diffuse component that encounters the laser beam wings.  By contrast, the cascaded collimator does not contain this diffuse component and therefore shows a less pronounced rate of growth.  This effect is well-captured by our three-dimensional theory, although the measured spectral HWHM was slightly lower than predicted, as will be discussed later.

\subsection{Optical pumping}
In addition to the saturation behavior explored earlier, we also deduced and quantified the influence of high probe laser intensity optical pumping on the hyperfine transitions. In Fig.~\ref{fig:hyperfine_levels}, we show broader spectra inclusive of all relevant hyperfine transitions as we scanned the laser frequency. The laser scan rate was approximately 12 GHz/second, and hence the laser frequency stayed within the natural linewidth $\Gamma$ = 6.1 MHz of any resonance for approximately 0.5 ms, longer than both the 9 $\mu$s transit time as well as the optical pumping time. Optical pumping to the dark $F = 1$ ground state can occur, whence less fluorescence is collected on the $F=2\rightarrow F'=1,2$ transitions. Once atoms are driven into the excited states labeled by $F'=1, 2$, they can spontaneously decay back to ground states label by $F=1, 2$ since they are allowed by the selection rule (indicted by the dashed lines). Atoms in $F=2$ state can by excited by the light field into the excited state again, but those pumped into $F=1$ are 6.8 GHz detuned away, hence they will not fluoresce anymore. The optical pumping rate to the dark state $F=1$ increases with laser intensity and thus relatively less fluorescence can be collected in comparison to the cycling $F=2$ to $F'=3$ transition. This effect is more pronounced at high saturation parameter, as shown in the spectra plotted in Fig.~\ref{fig:dualspectra}. This effect was also confirmed by multi-level master equation simulations that computed the excited state populations (and therefore the emission line strengths) for each hyperfine transition. We discuss these simulations in detail in section \ref{master_equation}. 

\subsection{Long term collimator output}
\begin{figure}[b]
\centering
\includegraphics[width = 0.45\textwidth]{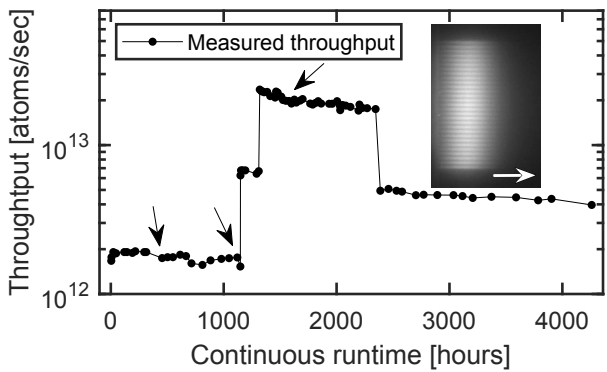}
\captionsetup{justification=centerlast}
\caption{
Atomic beam production continuous test.
Experimentally measured total throughput (error bars of 14\% not shown) for a micro-fabricated silicon collimator with 29 channels ($l \times w \times h$ = 3 $\si\mm$ $\times$ 0.1 $\si\mm$ $\times$ 0.1 $\si\mm$ each) versus running time plotted on a log scale. The oven was kept running in a continuous mode of operation. Temperature differences across the oven body of 5, 10 and 13 C correspond to oven operating temperature at 100 C, 125 C and 150 C. A camera image as an inset for the atomic beam fluorescence right after the channel exit indicates all 29 channels are open (no clogging). 3 black arrows mark the time at 454, 1120 and 1510 hours when we took camera images for the collimator output. The white arrow marks the propagation direction of the atomic beams.
}
\label{fig:throughput}
\end{figure}
While the collimating device is relatively simple, for practical applications such as clocks and other precision instruments, it needs to be run for years without servicing.  Traditional beam clocks have had to deal with the issue of nozzle clogging \cite{lucas2013atomic}.  For miniature, chip-scale collimators it is an open question to what extent clogging occurs in practice.  To address this issue, we performed limited lifetime tests under continuous operation for a period of 6 months at various fixed temperatures.  Our tests could both i) probe the overall flux emitted from the source using a photodetector to capture the fluorescence as well as ii) measure the output of every single individual microchannel using a CCD camera focused directly on the output of the chip.  Such channel-by-channel testing is not easily accomplished using 3-dimensional capillary array sources \cite{lucas2013atomic,Senaratne2015}, and our results are therefore a novel test of atomic beam behavior.  For our tests, the Rb oven body was kept at a constant temperature, while the chip temperature was between 10 and 30 degrees higher. For the total flux measurement, we probed the atomic beam 60 mm downstream from the chip after the beam had passed through a stainless steel plate with a 9 mm hole that was 36 mm away from the chip.  This aperture prevented any Rb vapor from accumulating in the probe region and contaminating the measurement.  As before, the laser probe was perpendicular to the atomic beam with a similar fluorescence collection, detection and amplification setup. The optical collection efficiency was 1.15\% and the power of the probe laser was 247 $\mu$w, with a saturation parameter $I/I_{sat} = 3.81$.
The spectrum of the fluorescence was deconvolved to obtain the transverse velocity distribution, which was used to calculate an average scattering rate $\overline{R_{sc}}$. We used $\overline{R_{sc}}$ to calibrate the total throughput of the collimator \cite{schioppo2012compact}. The result is shown in Fig.~\ref{fig:throughput}.

We set the oven temperature to 100 degrees at the beginning, 125 degrees at 1148 hours, 150 degrees at 1319 hours and 125 degrees again at 2387 hours. 
At the time of writing, this oven had run for over 4200 hours and none of the microchannels showed any sign of clogging. Pictures of the atomic beams near the nozzle were taken several times during the test, and showed that all 29 channels remain unclogged.
It can be noticed in Fig.~\ref{fig:throughput} that the flux is decaying slowly, which might result from the migrating of Rb inside the oven from hotter spots to colder spots. Coating of Rb on the vacuum windows over time and the drift of laser alignment may also contribute to this decaying effect.  
Since the flux is around 11 times higher at 150 degrees and 3 times higher at 125 degrees, our result implies a continuous operation time at 100 degrees of over 19000 hours without failure. This test proved that our microfabricated atomic beams are reliable and robust at different temperatures and can have a very long lifespan.

\section{Theory}

\begin{figure*}
\begin{minipage}{0.35\textwidth}
\centering
\subfloat[]{\label{fig:angf}\includegraphics[width = 0.95\textwidth]{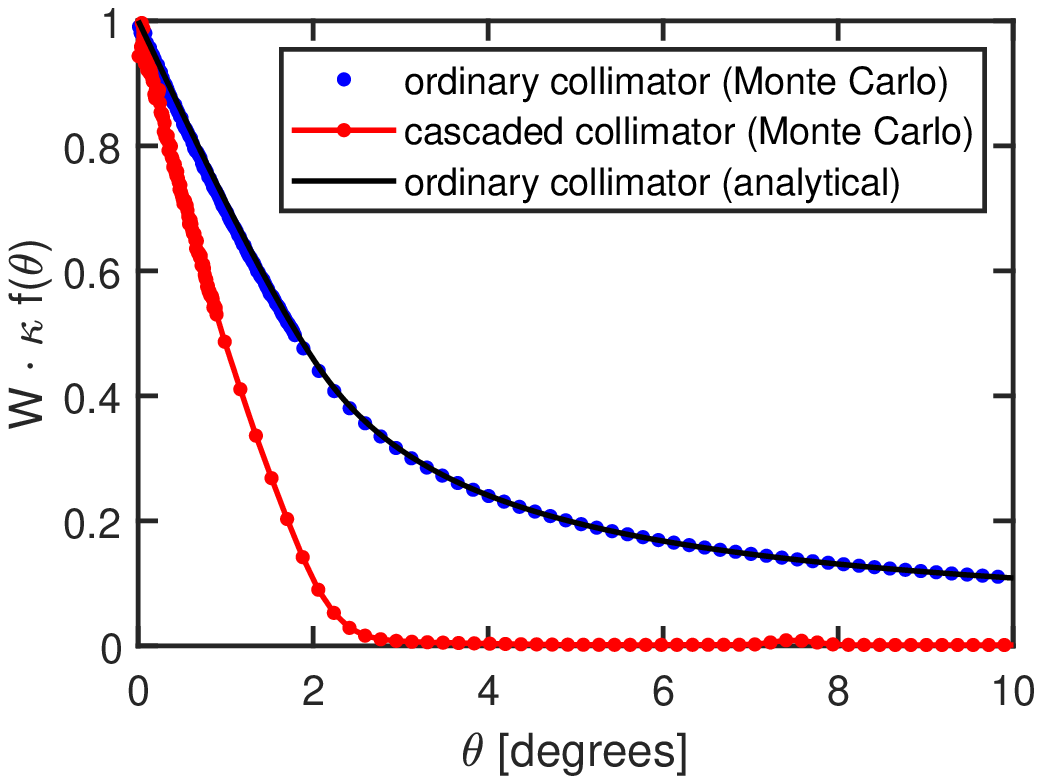}}
\end{minipage}
\begin{minipage}{0.43\textwidth}
\centering
\subfloat[]{\label{fig:imprate_intensity}\includegraphics[width = 0.95\textwidth]{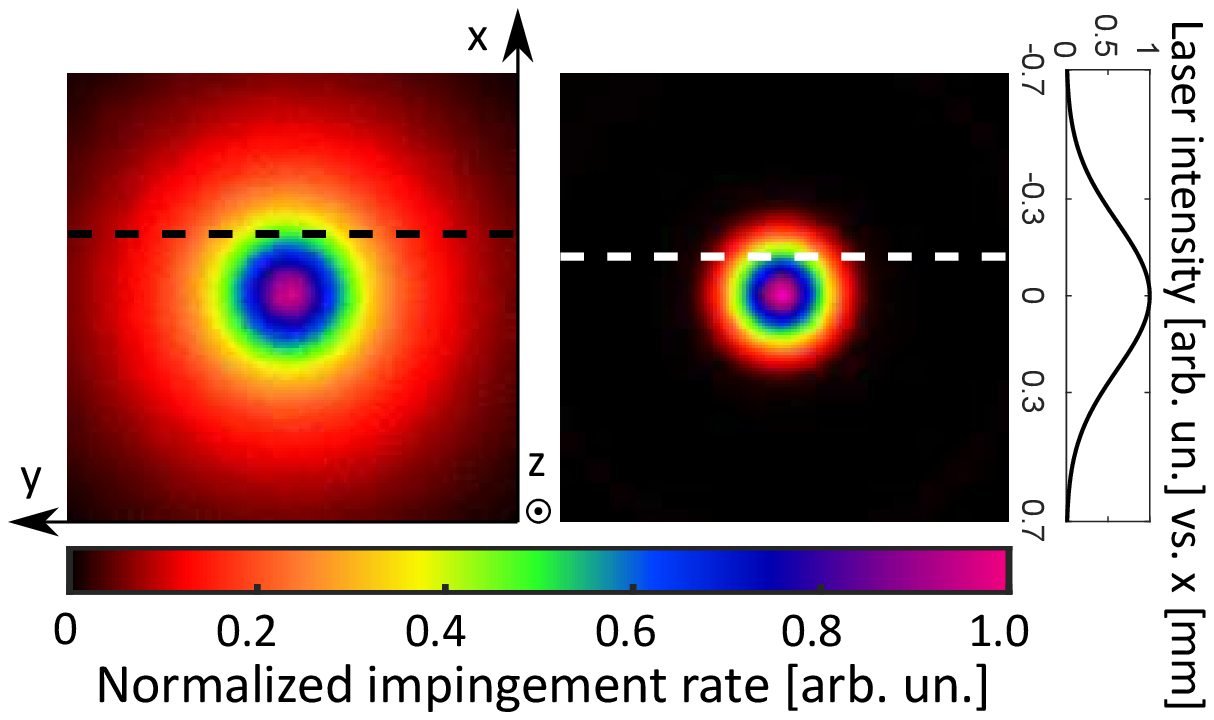}}
\end{minipage}
\begin{minipage}{0.37\textwidth}
\centering
\subfloat[]{\label{fig:spectra_off_x}\includegraphics[width = 0.95\textwidth]{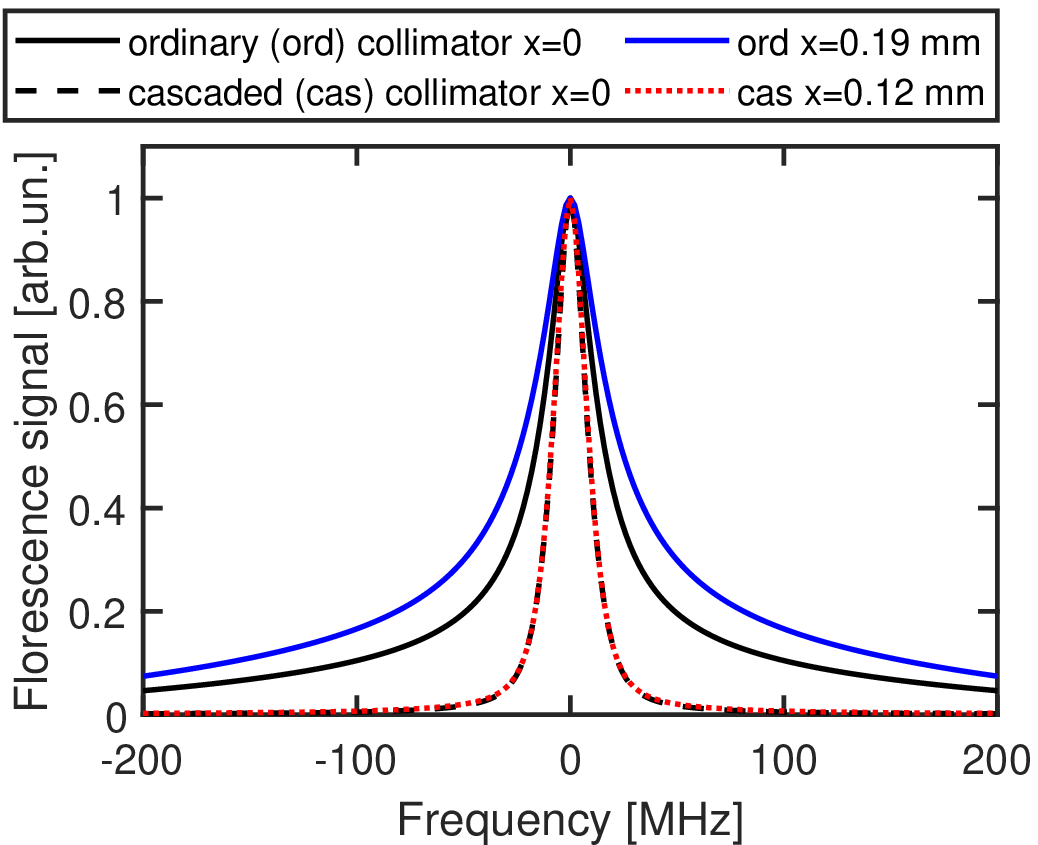}}
\end{minipage}
\begin{minipage}{0.36\textwidth}
\centering
\subfloat[]{\label{fig:spectra_off_z}\includegraphics[width = 0.95\textwidth]{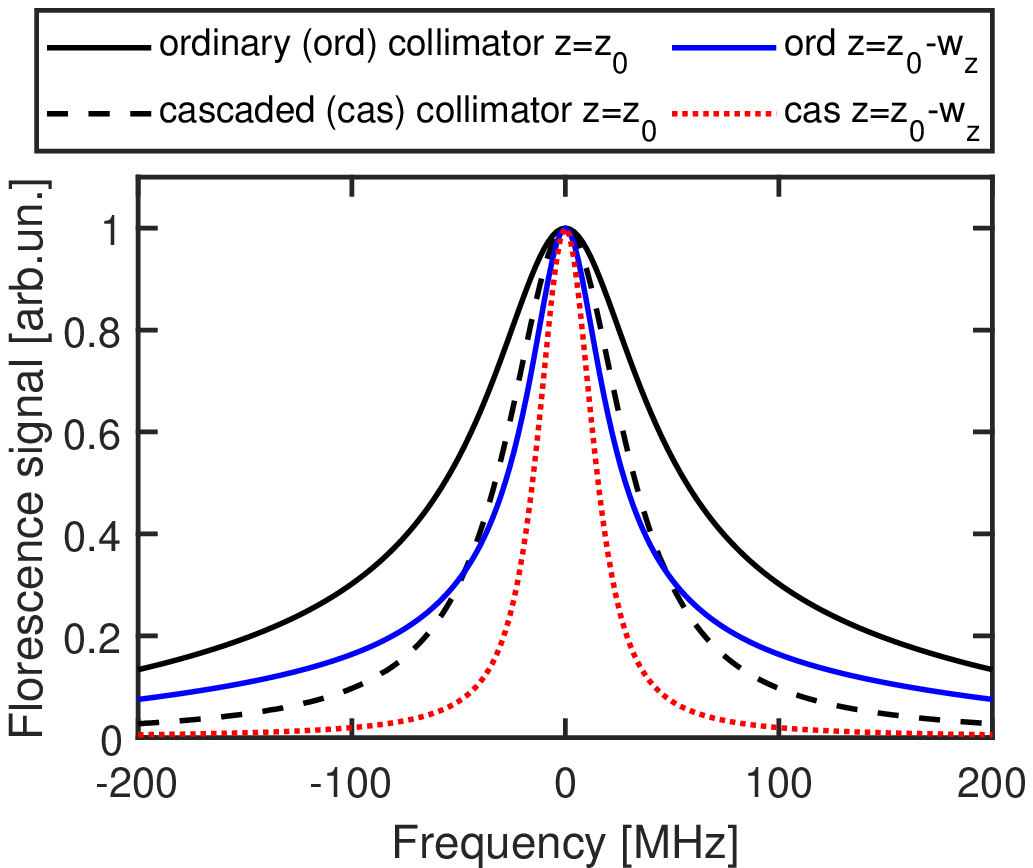}}
\end{minipage}
\captionsetup{justification=centerlast}
\caption{Monte-Carlo simulation of atomic beams allows full characterization of atomic fluorescence spectra.  
(a) The angular distribution functions $\kappa f(\theta)$ multiplied by their corresponding transmission probabilities (or Clausing factors) $W$. Blue and red are for Monte Carlo simulations. The black solid line represents the analytical result available only for the ordinary collimator \cite{Beijerinck1975,Steckelmacher1978}.
(b) The predicted impingement rate distribution for an ordinary collimator and a cascaded collimator. Each target is 1.4 mm by 1.4 mm in size with $10\:\si{\um} \times 10\:\si{\um}$ pixels. Impingement rates are, respectively, normalized to their peak values for these two targets. For reference, the laser beam intensity profile in the target plane is plotted to the right side. Circles sharing the same impingement rate form contours. The dashed line is tangent to a contour, where the impingement rate has dropped to a half. 
\space\space The computed fluorescence spectra for the ${}^{87}$Rb D$_2$ $F=2$ to $F'=3$ transition show the atomic/laser beam inhomogeneous effects.
(c) At low saturation parameter $s_0=1$, the computed spectrum is broader at $x=0.19$ for the ordinary collimator because of the diffuse component of the atomic beam. This component is missing in the cascaded collimator, hence the spectra are very similar at $x=0$ and $x=0.12$. The black (white) dotted lines in (b) indicate x=0.19 (0.12).
(d) At high saturation parameter $s_0=100$, the computed spectrum is narrower in the Gaussian wings compared with the center. Shown are spectra along different single lines on the $y-O-z$ plane, where the laser intensity is at its peak value ($z=z_0$) or at $1/e^2$ of its peak value ($z=z_0-w_z$).}
\label{fig:angf_imp_offx}
\end{figure*}

Our theoretical calculations combined i) Monte-Carlo simulations of the atomic flux in the molecular flow regime based upon the actual experimental geometry of the collimator imported from a CAD model and ii) an atom-by-atom computation of the fluorescence spectrum based upon each atom's interaction with the laser beam along its particular trajectory using the geometry defined in Fig.~\ref{fig:oven}.  For the latter, we used a master equation simulation to compute the population of each excited hyperfine level to deduce the fluorescence rate.  Details of these calculations are provided in the subsequent sections. 

\subsection{Monte Carlo simulations}
We used Molflow+, a test-particle Monte Carlo simulator dealing with molecular flow, to simulate atomic trajectories that pass through the laser beam \cite{ady2014molflow+}.  We assume that the collimators operate largely in the transparent regime \cite{giordmaine1960molecular}, where the mean free path $\lambda$ for atomic collisions is much larger than the length of an individual collimating channel $l_c$.  For our parameters, $l_c = 3$mm, while $\lambda$ can be estimated using the following equation  
\begin{equation}
    \lambda=\frac{k_B T}{\sqrt{2}\pi d^2 P}
\end{equation}
where $P$ is the pressure and $d$ the atomic diameter.  This condition begins to break down at temperatures of 120 C (150 C) for the single-stage (cascaded) collimators, as shown in Table \ref{tab:mfp}.  However, uncertainties in the exact temperature and atomic cross-section make this an approximate assessment.  In practice, we have not seen any discernable deviation from the transparent regime for oven temperatures up to 150 C for both collimators.

\begin{table}[b]
\captionsetup{format=plain, justification=centerlast}	
\caption{\label{tab:mfp}%
A table shows the rubidium vapor pressure $P$ and estimated mean free path $\lambda$ versus different temperatures $T$.  The vapor pressure is computed based on fits from Ref. \cite{alcock1984vapour} (Ref. \cite{yaws1995handbook}). The effective atomic diameter $d$ can be estimated from the Rb-Rb collision cross section $\sigma=1397$ $\AA ^2$ (i.e. $\sigma=\pi d^2$). 
}
\begin{ruledtabular}
\begin{tabular}{ccc}
\textrm{$T$ [C]}&
\textrm{$P$ [Pa]}&
\textrm{$\lambda$ [mm]}\\
\colrule
50 & 6.5(4.7)E-4 & 345.0(479.7) \\
70 & 3.5(2.6)E-3 & 68.4(90.5) \\
90 & 1.6(1.2)E-2 & 16.3(20.7) \\
110 & 5.9(4.8)E-2 & 4.5(5.6) \\
130 & 2.0(1.6)E-1 & 1.4(1.7) \\
150 & 5.9(4.9)E-1 & 0.5(0.6) \\
\end{tabular}
\end{ruledtabular}
\end{table}

The output beam can be fully characterized by an angular distribution function $\kappa f(\theta)$:
\begin{equation}
	d^3\dot{N}=(\dot{N}/\pi) \kappa f(\theta) d^2\Omega \Gamma_{\theta}(v)F(v)dv,
\end{equation}
where we use the notation of \cite{Beijerinck1975}. For a single orifice, $f(\theta) = \cos{\theta}$ and therefore $\kappa = 1$.  More generally for a tube, $\kappa \sim 1/W$ is the ``peaking factor'' of the collimator, depending inversely on its transmission probability $W=(4d/3l_c)(1+4d/3l_c)^{-1}$, where $l_c,d$ are the tube length and diameter, respectively \cite{Beijerinck1975}. 

One can import CAD geometries for their collimators into Molflow+ and specify the inlet surface as an cosine emitter with sticking factor=1 mimicking the physics that atoms can return to the source. Sticking factors for all internal surfaces of the collimator are set to be zero since no Rb chemical absorption/physical condensation is assumed for crystalline silicon maintained at a temperature that is higher than the rubidium vapor temperature. At the channel exit, another facet with sticking factor=1 captures all atoms emitted and records their angles while hitting the surface. The atomic gas is considered to be isothermal neglecting the thermalization over the whole simulation when atoms enter the microchannels kept at higher temperature than the oven. After releasing enough test particles since the statistical error varies with $1/\sqrt{N}$ \cite{ady2016monte}, the angular distribution function can be recovered using the following relation without distinguishing their speed distribution
\begin{equation}
	 C(\theta_{n}) \propto \int_{\theta_{i}}^{\theta_{f}} \int_{0}^{2 \pi} \kappa f(\theta_n) d^2 \Omega,
\end{equation}
where 
\begin{eqnarray} 
	&& d^2 \Omega = sin \theta d\theta d\phi \nonumber, \\
	&& \theta_{i}=\theta_{n}-\Delta \theta/2 \nonumber, \\
	&& \theta_{f}=\theta_{n}+\Delta \theta/2 \nonumber, \\
\end{eqnarray}
with $C(\theta_{n})$ represents the total number of particle falling into the $n_{th}$ bin spanned by the differential solid angle integrated over the specified range and $\Delta \theta$ represents the $\theta$ angle sampling step. We did not sample $\phi$ angle assuming the angular distribution function has a built-in rotational symmetry with respect to its center line ($z-$axis), given that the cross-sectional shape of collimating tubes has negligible influences to its angular distribution function based on Ref. \cite{Steckelmacher1978}. $\kappa f(\theta)$ can then get appropriately normalized through
\begin{equation}
	 \int_{0}^{\pi/2} \int_{0}^{2 \pi} \kappa f(\theta) d^2\Omega = \pi.
\end{equation}
Using Molflow+'s advanced facet parameters, we have finer samplings imposed for small angles $\theta$ to achieve better precision depicting the peaking behavior of the angular distribution function: $\Delta \theta =$ 3.14E-4 for $ \theta \in [0,3.14E-2]$ for the single-stage collimator while $\Delta \theta =$ 1.57E-4 for $ \theta \in [0,1.57E-2]$ for the cascaded collimator. Another 500 values are equally sampled for $\theta$ angle outside these ranges up to $\pi/2$, and only 1 value for $\phi$ angle from $0$ to $2\pi$. 

Results for the single-stage collimator and the cascaded collimator are plotted in Fig.~\ref{fig:angf}, where the angular distribution functions have been multiplied by its corresponding transmission probability for comparison with the same peak value 1 noticing the fact that $W\cdot \kappa f(\theta=0)=1$. For simple geometries as the ordinary type of collimators, an analytical expression for the angular distribution function $\kappa f(\theta)$ exists, hence we plot the results given by Molflow+ together with the analytical expression. The good agreement between the two confirms the validity of Molflow+, which can be applied to any non-trivial structures, e.g. the cascaded collimators.

At the center axis of the Gaussian laser beam as shown in Fig.~\ref{fig:oven}, we then set up two square targets facing towards the collimator output. The two targets are sitting 6 $\si{\mm}$ away from the nozzle exit and detecting the number of hits per unit area per unit time. The impingement rate distributions for two different types of collimators are shown in Fig.~\ref{fig:imprate_intensity}. More than $99\%$ of atoms are actually emitted into the halos for the traditional type single-stage collimator. This can smear out the fine spectra features, contaminate nano structures or micro cavities \cite{keith1991interferometer,gregoire2016static,Sukenik1993}, and reduce transparency of optical accesses. In contrast, the cascaded collimator with the identical physical size can produce purer atomic beams with 40 times fewer atoms emitted into large angles while maintaining the same axial beam intensity of $nA\bar{v}/(4\pi)$ \cite{lucas2013atomic}.

\subsection{Fluorescence spectral computation}
Under the transparent regime, the on-axis atom number density at a certain location $z$ with respect to the nozzle exit is given by 
\begin{equation}
    n(z)=\frac{n_0A}{4 \pi z^2},
\end{equation}
where $n_0$ is the atom number density inside the oven, and $A$ is the cross-section area of a single tube. For $N_t$=20 tubes with diameter $d$, the optical depth at $T$=100 $\degree$C can be estimated using
\begin{equation}
    OD=N_t n(z)\sigma d,
\end{equation}
where $\sigma=3\lambda^2/2\pi$ is the photon scattering cross section \cite{SteckRb2015}. OD is about 6\% at maximum for the ${}^{87}$Rb D$_2$ $F=2$ to $F'=3$ transition.
Neglecting the absorption, we can assume the laser beam intensity only varies with respect to $x$ and $z$ while it does not change much along its propagation axis y. The Gaussian laser beam intensity distribution can be written as
\begin{equation}\label{eq:laser_int}
    I=I_0 e^{-2(x^2/w_x^2+(z-z_0)^2/w_z^2)}.
\end{equation}
For multiple atomic beams, the measured output voltage is
\begin{eqnarray} \label{eq:fluorescence}
	&&V_f(\delta) = \hbar\omega_{0}R_{resp}G\eta A_b f_p N_t f_c^{-1} \dot{N}\int_{0}^{\infty} \dd{v} \int_{B} \dd{V} \times  \nonumber \\
	&& \frac{\displaystyle  \kappa f(\theta)F(v)}{\pi \displaystyle v r^2} R_{sc}(s(x,y,z), \delta-kv\sin{\theta}\sin{\phi}),
\end{eqnarray}
where $\delta$ is the laser detuning, $\hbar\omega_0$ gives the energy per photon, $R_{resp}$ is the responsivity of the photodiode, $G$ is the current amplifier gain, $\eta$ is the overall photon collecting efficiency, $A_b$ is the natural abundance of ${}^{87}$Rb, $f_p$ is the population fraction for a certain hyperfine level, $N_t$ is the total number of microcapillaries, $f_c$ is the correction factor \cite{Li2019} for the theoretical total throughput $\dot{N}$ and $B$ is determined by the fluorescence collecting volume. Numerical values of these scaling factors are given in Table \ref{tab:scaling_factor}. While seemingly complicated, Eq. (\ref{eq:fluorescence}) can be formulated into a simple linear algebra problem after discretizing the integral (see the appendix)
\begin{equation}\label{eq:format_in_text}
    \vec{V_f}=A(s_0)\vec{f},
\end{equation}
where $A(s_0)$ is a matrix mapping the angular distribution function $\vec{f}$ into the predicted fluorescence spectra $\vec{V_f}$ for a certain saturation parameter $s_0$. Theoretical results for Fig.~\ref{fig:theory_exp}(a)-(c) are readily obtained once $A(s_0)$ is computed for the ten different experimental saturation parameters. One can either predict the angular distribution function using Monte Carlo simulations and then predict the fluorescence spectra treating Eq. (\ref{eq:format_in_text}) as a simple forward problem, or measure the fluorescence spectrum and inversely find the angular distribution function \cite{gill1981practical,hansen2007regularization}.

In the low saturation parameter regime, a broader spectrum can be measured for the single-stage collimator while probing the fluorescence spectrum off the $y-O-z$ plane. We quantitatively verify this by computing the fluorescence spectra using Eq. (\ref{eq:fluorescence}) along the dashed lines and the center lines, and normalize their peak value to be 1 for convenience (see Fig.~\ref{fig:spectra_off_x}). That's why the 1-D approximation predicts the narrower spectrum compared to the measured data as reported in our previous work \cite{Li2019}. For the cascaded collimator, spectra measured at different $x-$locations are similar, since the number of atoms emitted from tube walls has already been suppressed by 40 times. Atoms passing through the region near the dotted lines for the cascaded collimator mainly come from the beam component rather than the diffuse component.

In the high saturation parameter regime, a narrow spectrum can be measured in the Gaussian wings of the laser beam compared to the center when probing the fluorescence spectrum on the $y-O-z$ plane. Following the same procedure above, the spectra are calculated along lines at different $z-$locations (see Fig.~\ref{fig:spectra_off_z}). The spectra computed at $z=z_0$ is broader than the spectra computed at $z=z_0-w_z$ where $s=s_0/e^2$ solely because the power broadening is larger. These spectra share the same non-trivial transverse speed distribution, and thus we can compare the power broadening directly. At the Gaussian wings of the laser beam, the spectral half width converges towards their intrinsic Doppler broadening as shown in Fig.~\ref{fig:spectra_off_x}. Analytical estimates are not available for the half width of the measured fluorescence spectra when the Doppler and power broadening are comparable to each other, considering the atomic/laser beam inhomogeneous effects if probing close to the source.

\subsection{Master equation simulations}\label{master_equation}
Fig.~\ref{fig:dualspectra} reveals that the red side of the fluorescence spectra manifests different line strength for the $F=2$ to $F'=2,1$ transitions under different laser intensities due to optical pumping effects. A closed-form analytical expression for $R_{sc}$, as required in Eq. (\ref{eq:fluorescence}), isn't available for these two transitions in contrast to the $F=2$ to $F'=3$ cycling transition. Therefore, to quantify these optical pumping effects, we need to run master equation simulations to track the dynamical evolution of the excited state populations contributing to the fluorescence.

\begin{figure}[htbp]
\begin{minipage}{0.45\textwidth}
\centering
\subfloat[]{\label{fig:2to3}\includegraphics[width = 0.9\textwidth]{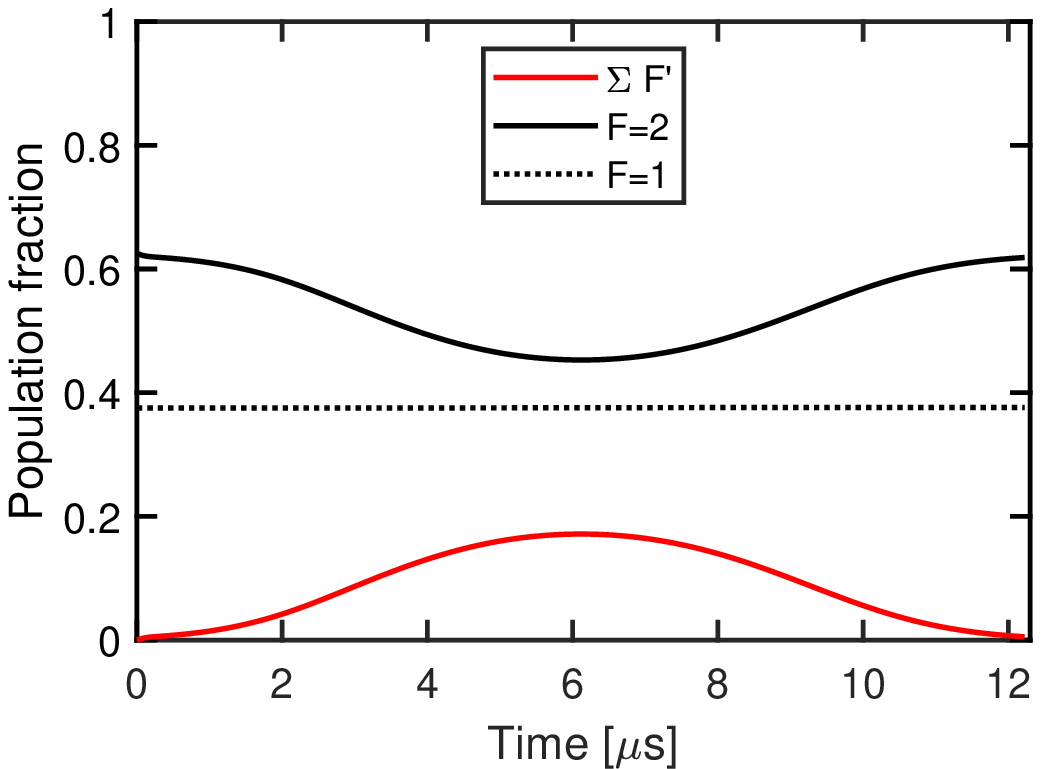}}
\end{minipage}
\begin{minipage}{0.45\textwidth}
\centering
\subfloat[]{\label{fig:2to2}\includegraphics[width = 0.9\textwidth]{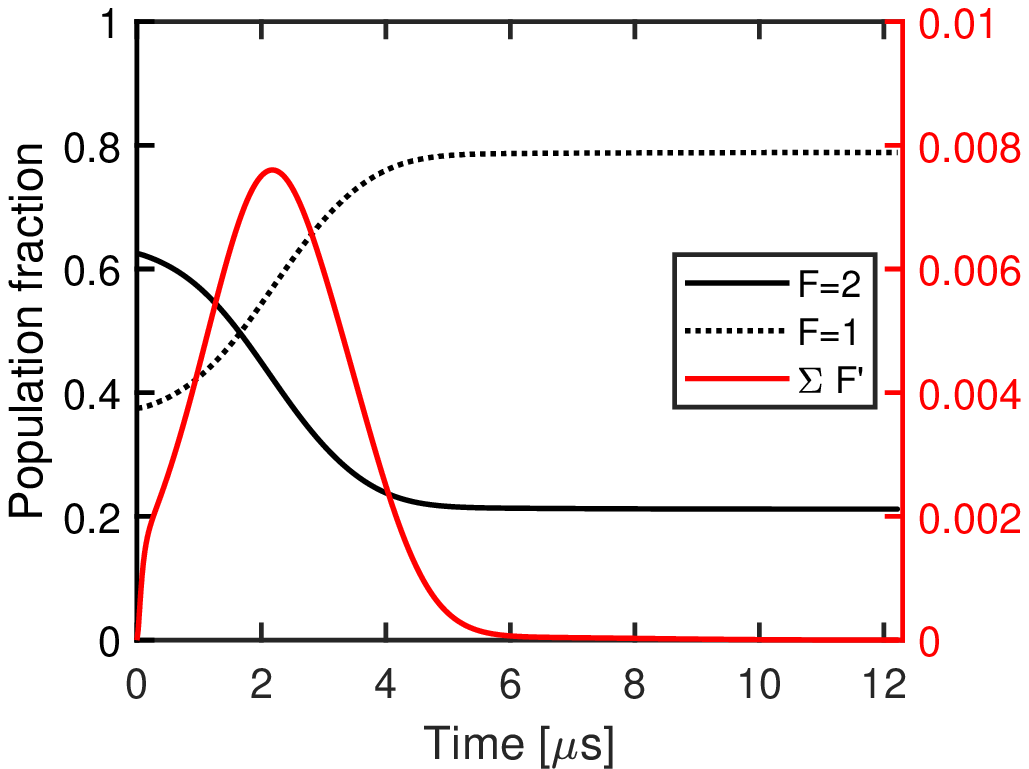}}
\end{minipage}
\captionsetup{justification=centerlast}
\caption{Master equation simulations. An ensemble of atoms are traveling through the Gaussian light field. The black curves show the time-evolution of the population in individual ground states. The red curve shows the time-evolution of the summed population in all excited states ($F'=0,1,2,3$).  The laser frequency is on-resonance with the ${}^{87}$Rb D$_2$ $F=2$ to $F'=3$ cycling transition for (a), and is on-resonance with the $F=2$ to $F'=2$ non-cycling transition for (b), where we have a separate axis on the right for the population in excited states.}
\label{fig:pumping}
\end{figure}

The master equation reads as the following \cite{steck2007quantum,Li2019}
\begin{equation}\label{eq:mastereq}
		\dot{\rho}(t) = -\frac{i}{\hbar}\comm{H(\bm{r},t)}{\rho(t)}+\Gamma \left ( \frac{2J'+1}{2J+1} \right ) \sum_{q=-1}^{1}\mathcal{D}[\Sigma_q]\rho(t),
\end{equation}
where the first part is for the atom-light interaction and the second part is for the spontaneous decay. At low light level (peak intensity $I_0$=3.8 mW/cm$^2$) and oven temperature T=100 C, an ensemble of atoms are assumed to be traveling with the most probable velocity $v_{pB}=327$ m/s for an atomic beam \cite{ramsey1956molecular} going through the light field corresponding to Eq. (\ref{eq:laser_int}), from ($z_0$-$1.5w_z$) to ($z_0$+$1.5w_z$) where the light intensity has dropped to a significantly low level . We initialize the density matrix according to the Boltzmann distribution.

Fig.~\ref{fig:pumping} shows the population dynamics for both $F'=3$ and $F'=2$ excitations. Our principal observation is that the $F'=3$ excitation can be effectively treated as two-level. Fig.~\ref{fig:2to3} shows that the excited state population tracks its steady state value 
\begin{equation}\label{eq:excited_pop_f3}
    P_e(F'=3,\vec{r}(t))=\frac{1}{2}\frac{I(\vec{r}(t))/I_{sat}}{1+I(\vec{r}(t))/I_{sat}}\frac{5}{8},
\end{equation}
where $I(\vec{r}(t))$ is the intensity at the instantaneous location $\vec{r}(t)$ of the atoms. The difference between Eq. (\ref{eq:excited_pop_f3}) and the master equation simulation was negligible. From Eq. (\ref{eq:excited_pop_f3}) we can get the on-resonance photon scattering rate $R_{sc}(\vec{r}(t))=\Gamma P_e(F'=3,\vec{r}(t))$ for atoms at any specific locations. 
Fig.~\ref{fig:2to2} shows the populations in $\ket{F=2}$, which can contribute to the collected fluorescence through temporarily occupying $\ket{F'=2}$ and are eventually transferred into the dark state $\ket{F=1}$ that has a 6.8 GHz detuning from the excitation field. An additional dark state is $\ket{F=2,m_F=0}$ for which there is a zero coupling matrix element to $\ket{F'= 2,m_F=0}$ \cite{foot2005atomic}. These states become substantially populated within $\sim$ 5 $\mu$s. The presence of a kink feature in the solid red curve at the beginning stage of the simulation is because the Rabi flopping takes a time about 1/$\Gamma$ to reach steady state due to spontaneous emission, and during that time the system evolves coherently. The integrated number of photons scattered is computed from the excited state probabilities and yields good quantitative agreement with the measured spectra in Fig.~\ref{fig:dualspectra}. These procedures were previously documented (see Ref. \cite{Li2019}) but not the master equation simulation data of Figure 6.

\begin{table*}[htbp]
\captionsetup{format=plain, justification=centerlast}	
\caption{\label{tab:scaling_factor}%
A table shows the scaling coefficients, as given in Eq. (\ref{eq:fluorescence}), participating into the fluorescence signal calculations.
}
\begin{ruledtabular}
\begin{tabular}{ccc}
\textrm{Symbol}&
\textrm{Meaning}&
\textrm{Value}\\
\colrule
$\omega_{0}$ & Laser frequency & 2$\pi\cdot$384.2 THz\\
$R_{resp}$ & Photodiode responsivity & 0.6 A/W \\
$G$ & Current amplifier gain & $1\times10^8$ V/A\\
$\eta$ & Photon  collecting  efficiency & $1.2\%$ \\
$A_b$ & Natural abundance & $27.8\%$ \\
$f_p$ & Fraction of population in $\ket{F=2}$ & 5/8 \\
$N_t$ & Number of tubes & 20 \\
$f_c$ & Correction factor & 2.4 (cascaded), 2.8 (ordinary) \cite{Li2019}\\
$\dot{N}$ & Total throughput at T=100 C & $4.2\times10^9$ s$^{-1}$ (cascaded), $1.7\times10^{11}$ s$^{-1}$ (ordinary)\\
\end{tabular}
\end{ruledtabular}
\end{table*}

\section{Conclusion}
In summary, we have developed NSFS for scenarios where laser intensities, atoms' spatial distributions and speed distributions are all wound together. The clogging-free continuous operation and the individual-channel addressability identify these on-chip elements as promising candidates, generating well-collimated atomic beams for clocks, interferometery or the atomic-optical device hybridization. \cite{elgin2019cold,hemmer1993semiconductor,brand2015atomically,liron2019chip,alaeian2019cavity}. Looking forward, combining Monte Carlo and master equation simulations allows the tracking of atom trajectories and their internal states simultaneously, which is indispensable for customizing the laser manipulation of atoms close to or within a chip-scale source itself \cite{metcalf2017colloquium}. The simple recipe phrased in linear algebra for reconstructing the speed/angular distribution of the atomic beams is important for precision spectroscopy \cite{zheng2019light} or atom scattering experiments \cite{sekiguchi2018scattering}, and useful for guiding collimator and oven designs \cite{wouters2016design,sharma2019atomic,xiao2019shaped}. 

\section*{Acknowledgments}
\addcontentsline{toc}{section}{\protect\numberline{}Acknowledgments}
We thank Scott Elliott, Nathan Mauldin and Frank Murdock at the MRDC Machine Shop for manufacturing oven test assemblies. We thank Richard Shafer at the IEN Laser Lab for the guidance on Optec femtosecond laser micro-machining system. C.L. acknowledges helpful discussions with Kevin Driscoll, Xin Xing and Shangguo Zhu.

\section*{Appendix}
\addcontentsline{toc}{section}{\protect\numberline{}Appendix}

Following Eq. (\ref{eq:fluorescence}), insert 
\begin{equation}\label{eq:dV}
    dV=r^2sin\theta dr d\theta d\phi
\end{equation}
and 
\begin{equation}\label{eq:Fv}
    F(v) \propto v^3 \cdot e^{-(v/\alpha)^2}
\end{equation}
into the expression for the detected fluorescence, and we get
\begin{equation}\label{eq:fluorescencevsdelta}
    V_f(\delta) \propto \int_{0}^{\infty} dv \int dr \int d\phi \int d\theta A(v,r,\theta,\phi,\delta,s_0)\kappa f(\theta),
\end{equation} 
where we've dropped all constants as coefficients and $A(v,r,\theta,\phi,\delta)$ represents 
\begin{equation}\label{eq:matrixA}
	A(v,r,\theta,\phi,\delta,s_0)=\sin \theta \cdot R_{sc}(r,\theta,\phi,s_0,\delta,v) \cdot v^2 e^{-(v/\alpha)^2}.
\end{equation} 
The field of view of our imaging system defines the collecting volume $B$ and thus determines the integral lower and upper limit for $r, \theta, \phi$ in Equation \ref{eq:fluorescencevsdelta}. If we define
\begin{equation}\label{eq:hdeltatheta}
	A(\delta,\theta,s_0)=\int_{0}^{\infty} dv \int dr \int d\phi A(v,r,\theta,\phi,\delta,s_0),
\end{equation} 
then Equation \ref{eq:fluorescencevsdelta} can be re-written as the following simple expression
\begin{equation}\label{eq:simplifiedfluorescence}
	V_f(\delta) \propto \int d\theta A(\delta,\theta,s_0) \kappa f(\theta).
\end{equation} 
The discrete version of Equation \ref{eq:simplifiedfluorescence} will be 
\begin{equation}\label{eq:discretefluorescence}
	V_f(\delta_i) \propto \sum_{j}  \Delta \theta A(\delta_{i},\theta_{j},s_0) f(\theta_{j}).
\end{equation} 
Now, the theoretical prediction for the fluorescence spectrum has been completely formulated into a problem of the following type
\begin{equation}\label{eq:problemtype}
	\vec{V_f}=A(s_0)\vec{f},
\end{equation} 
where $\vec{f}$ represents the angular distribution function $f(\theta)$, $\vec{V_f}$ represents the fluorescence spectrum $V_f(\delta)$ and matrix $A(s_0)$ represents the mapping between the two under different saturation parameters. Interestingly, matrix $A(s_0)$ does not depend on what type of collimator/angular distribution function we have at all. It intrinsically characterizes the mapping once the laser probe geometry and power are well defined. For absolute fluorescence signal calibration, we absorb previously dropped scaling coefficients back to Eq. (\ref{eq:problemtype}) by the end of the computation. Scaling factors defined in Eq. (\ref{eq:fluorescence}) are summarized in Table \ref{tab:scaling_factor}.


\begin{thebibliography}{42}
\bibitem [{\citenamefont {{Molflow+ is a Monte Carlo code developed at CERN by
  R. Kersevan and M. Ady}}()}]{ady2014molflow+}%
  \BibitemOpen
  \bibfield  {author} {\bibinfo {author} {\bibnamefont {{Molflow+ is a Monte
  Carlo code developed at CERN by R. Kersevan and M. Ady}}},\ }\href@noop {}
  {\bibinfo  {journal} {{website: cern.ch/molflow}}\ }\BibitemShut {NoStop}%
\end{thebibliography}
\end{document}